# Laminar flow of viscous fluid around elliptical contours under angle of attack

Alexander G. Petrov[a], Artem D. Sukhov[b]

[a] *Ishlinsky Institute for Problems in Mechanics RAS, Prospekt Vernadskogo, 101-1, Moscow, 119526, Russia, petrovipmech@gmail.com*
[b] *Moscow Institute of Physics and Technology (National Research University), Dolgoprudny, 141701, Russia, sukhov.ad@phystech.edu*

**Abstract**
The planar problem of a viscous laminar flow around elliptical cylinders under angle of attack is considered. From the solution of the laminar boundary layer equations using the Loytsyansky local similarity method, the shear stress at the ellipse boundary and the separation point is found. From the separation points velocities equality, the circulation is found. A complete solution to the problem of the velocity and pressure field outside the boundary layer is also constructed. The coefficients of lift and resistance are found depending on the angle of attack and the ellipse axes ratio. The theoretical results are compared with the available experimental data and direct numerical solutions of the Navier-Stokes equations.

**Keywords 1**
Ellipse, laminar flow, viscous fluid, boundary layer, lift force, resistance

## 1. Introduction

The investigation of the separated flow around cylinders by a planar laminar flow of a viscous fluid has been carried out since the beginning of the 20th century. For this, the Prandtl boundary layer theory is used, as well as similarity theory methods for the approximate solution of these equations. The monograph [6] provides a large overview of such investigations of various contours, including elliptical cylinders, separation points are calculated by the Karman-Palhausen method, they are carried out by German scientists. They assume that the separation point is recieved from the equality of the shear stress to zero, and the circulation is found from the equality of the velocities at the separation points. The development of the method and the calculations themselves were carried out by prominent German scientists: Karman, Pohlhausen, Howarth, Falkner, Skan, Dryden and other. Soviet scientists made a significant contribution to the theory of boundary layer methods: Loitsyanskii, Kotchin and other. To current date, there are a limited number of calculations and experimental data for particular cases. This investigation is devoted to the development of a semi-analytical method for the complete solution of the problem of flow around elliptical cylinders with an different ratio of the axes and at any small angles of attack.

## 2. Potential flow around the ellipse

We write the ellipse equation with semiaxes $a_{el}, b_{el}$ on the complex plane $z = x + iy$ in parametric form $z = a_{el}\cos t + ib_{el}\sin t$.

The flow around the ellipse under angle of attack $\theta$ with velocity $|v_\infty|e^{i\theta}$ and with circulation $\Gamma$ at infinity creates the following velocity distribution on ellipse boundary [1]

---

$$v(t) = \frac{\Gamma/(2\pi) - (a_{el} + b_{el})|v_\infty|\sin(t-\theta)}{q(t)},$$

$$q(t) = \frac{1}{2}|a_{el} + b_{el} - i(a_{el} - b_{el})(\cos 2t - \sin 2t)|.$$
(2.1)

For oblate ellipse $b_{el} \gg a_{el}$ velocity at the point $t = 0$

$$v(0) = \frac{\frac{\Gamma}{2\pi} + (a_{el} + b_{el})|v_\infty|\sin\theta}{b_{el}}$$

becomes large. To remove this feature, the circulation is chosen so that the numerator at this point turns to zero. This choice of circulation for profiles with a sharp edge is called the Chaplygin-Zhukovsky postulate. The force acting on the contour is perpendicular to the velocity and is called lift force. The circulation $\Gamma$ and the lift force Y are as follows

$$\Gamma = -2\pi(a_{el} + b_{el})|v_\infty|\sin\theta, \quad Y = 2\pi\rho(a_{el} + b_{el})|v_\infty|^2 \sin\theta.$$
(2.2)

With this choice of circulation, velocity on ellipse boundary (2.1) reduced to form

$$v(t) = \frac{a_{el} + b_{el}}{q(t)}(-\sin\theta - \sin(t-\theta))$$
(2.3)

At the critical point

$$t = t_{cr} = \pi + 2\theta$$
(2.4)

velocity is equal to zero.

## 3. Separated flow around an ellipse in a viscous fluid flow

In a fluid with a kinematic viscosity coefficient $\nu$ at high Reynolds number $Re = \frac{2av_\infty}{\nu}$, a boundary layer of thickness $\delta = 2a_{el}/\sqrt{Re}$ is created near the contour. In this layer, starting from the critical point to the separation point, the velocity to the right and to the left of the critical point changes from 0 to a value close to the potential flow velocity $v$. On the profile boundary creates a shear stress is equal in order of magnitude $\tau_0 = \mu v/\delta$. From a more complex analysis of the flow in the boundary layer using the approximate Loitsiansky method, the shear stress is determined by the formula

$$\tau = \frac{\mu[v(s)]^{1+b/2}\alpha(f+f_s)^\gamma}{\sqrt{va\int_0^s v(s')^{b-1}ds'}}, \quad \frac{1}{a}(f+f_s) = \frac{v'(s)}{v(s)^b}\int_0^s v(\xi)^{b-1}d\xi + \frac{f_s}{a},$$

where $a = 0,44; b = 5,75; \alpha = 1.3; f_s/a = 0.202$ - approximation constants.

Hence, for the field of unseparated flow $\tau > 0$ the inequality $f + f_s > 0$ is obtained.

The critical points are determined from the condition that the shear stress is zero $\tau$, that is, from the follows equations

$$\frac{v_r'(s)}{[v_r(s)]^b}\int_0^s v_r(\xi)^{b-1}d\xi + 0.202 = 0, \quad \frac{v_l'(s)}{[v_l(s)]^b}\int_0^s v_l(\xi)^{b-1}d\xi + 0.202 = 0.$$

Where the integration is over the arc length, which is measured from the critical point. The same equation for determining the separation point is given in the monograph [4] in the page 2008.

The critical point (2.4) is displaced, when the ellipse is flown around under angle of attack with the separation points

$$t = t_{cr}(\Delta t) = \pi + 2\theta + \Delta t \qquad (3.1)$$

Through the parameter $\Delta t$ can be expressed

$$\Gamma(\Delta t) = -2\pi(a_{el} + b_{el})|v_\infty|\sin(\theta + \Delta t), \quad Y = 2\pi\rho(a_{el} + b_{el})|v_\infty|^2 \sin(\theta + \Delta t). \qquad (3.2)$$

Instead of finding the circulation, the value of the parameter $\Delta t$ is found.

Let us present the analytical dependences, with the help of which it is easier to calculate the separation points.

Velocity at the ellipse boundary and arc element

$$\sigma(t) = \frac{ds}{dt} = |-a_{el}\sin t + ib_{el}\cos t|$$

When $t < t_{cr}(\Delta t)$ the velocity $v_r$ and the arc element $\sigma_r$ depends on $t_r = t_{cr}(\Delta t) - t$ and on the parameter of the critical point $\Delta t$.

$$v_r(t_r, \Delta t) = v(t_{cr}(\Delta t) - t_r), \quad \sigma_r(t_r, \Delta t) = v(t_{cr}(\Delta t) - t_r).$$

Velocity $v_r$ at the $t_r = 0$ equal to zero and at the $t_r > 0$ remains positive up to the separation point $t_{r0}$.

Separation point $t_{r0}$ is found from the equation

$$\frac{v'_r(t_r, \Delta t)}{v_r(t_r, \Delta t)^b \sigma(t_r)} \int_0^{t_r} v_r(\xi, \Delta t)^{b-1} \sigma_r(\xi, \Delta t) d\xi + 0.202 = 0. \qquad (3.3)$$

The separation point $t_{l0}$ is defined similarly at the $t > t_{cr}$.

$$\frac{v'_l(t_l, \Delta t)}{v_r(t_l, \Delta t)^b \sigma(t_l, \Delta t)} \int_0^{t_l} v_l(\xi, \Delta t)^{b-1} \sigma_l(\xi, \Delta t) d\xi + 0.202 = 0, \qquad (3.4)$$

where $t_l = t - t_{cr}(\Delta t)$

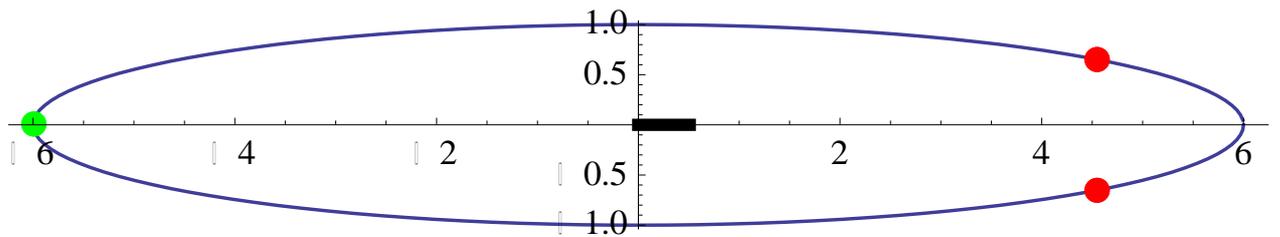

Fig.1. Symmetrical flow around the ellipse.

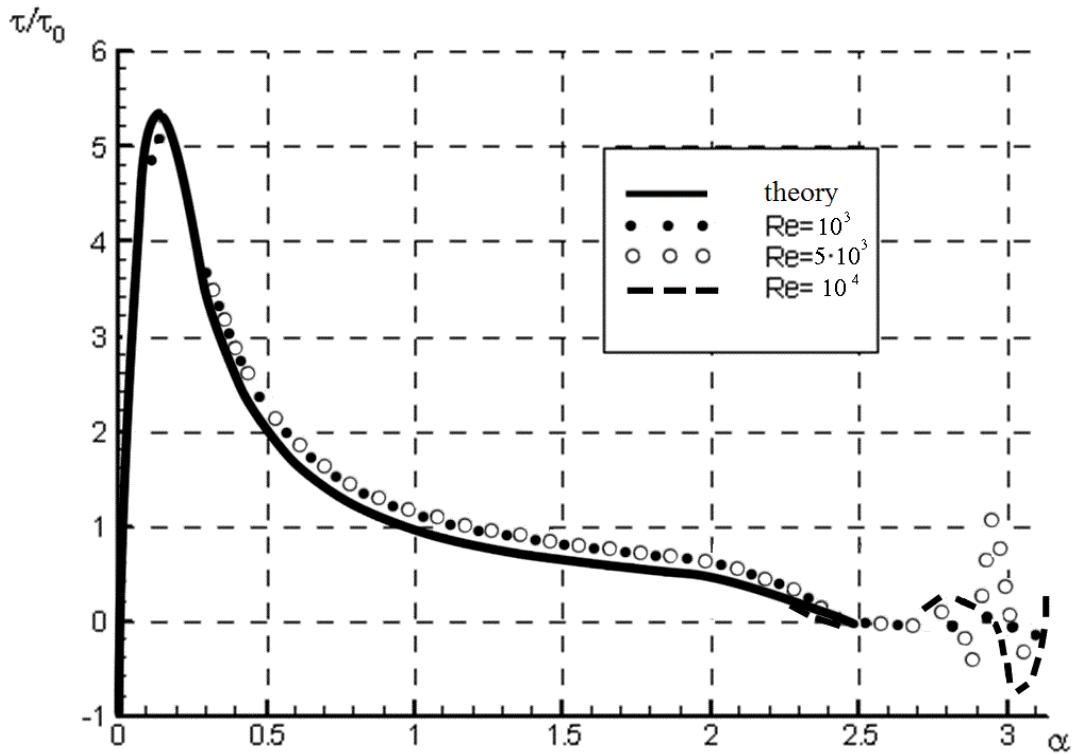

Fig. 2. Comparison with direct numerical solution of the Navier-Stokes equations for different Reynolds numbers

If the angle of attack is zero $\theta = 0$, then the flow is symmetric about the semi-axis of the ellipse $a_{el}$ and the circulation is zero, also $\Delta t = 0$. In this case, the solution about the flow around the ellipse is constructed to the end, assuming that the pressure in the separation zone is constant. In the fig. 1 shows the critical point (green) and separation points (red) for a symmetric flow around an ellipse with semiaxes $a_{el} = 6$, $b_{el} = 1$. In the critical point $t_r = 0$ and in the separation point $t_{r0} = 2.43$. On the ellipse, the separation point has coordinates $x_{r0} = a_{el} \cos t_{r0} = 4.54$, $y_{r0} = b_{el} \sin t_{r0} = 0.65$.

In fig. 2. Comparison of the shear stress with the direct numerical solution of the Navier-Stokes equations at $Re = 10^3; 5 \times 10^3; 10^4$, carried out in [5], is presented.

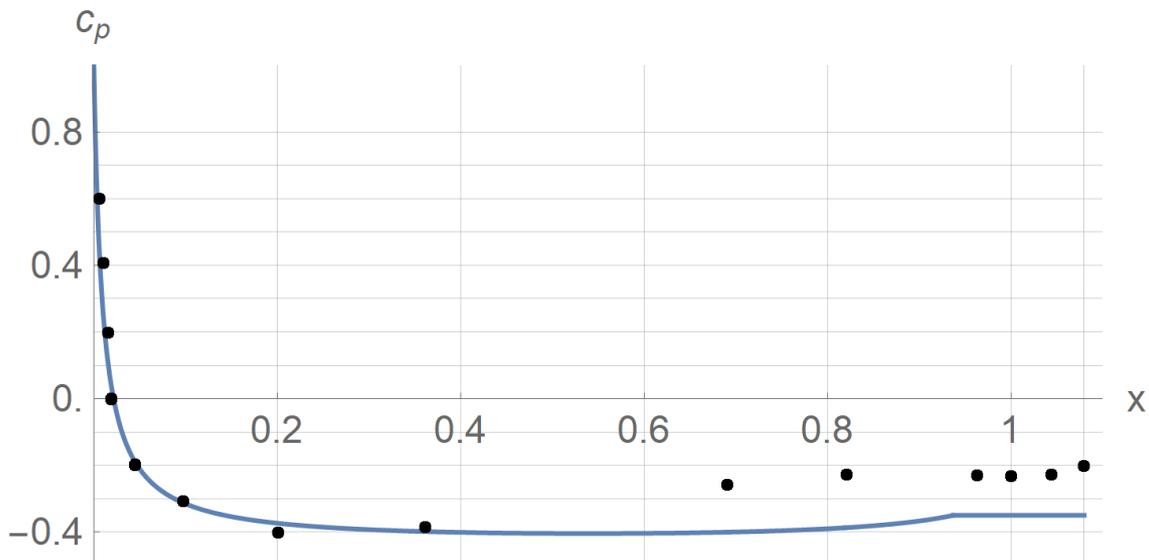

Fig. 3. Comparison with experiment

Calculation of pressure coefficient for flow around ellipse with axis $a_{el} = 0.54, b_{el} = 0.1$ was done. In fig. 3. the comparison with experiment [6] (p. 479, black points), theoretical dependence of the pressure coefficient $c_p(t) = 1 - v(t, \Delta t)^2 / v_\infty^2$, $\Delta t = 0$ on the distance to the critical point along the axis of the ellipse $x = 0.54(1 - \cos\alpha)$. Reproduced dependence $c_p(\pi - \alpha)$ depending on x. Velocity $v(t, \Delta t)$ is calculated by (2.1) with zero circulation to the critical point with the parameter $\alpha_{right} = \pi - t_{right} = 2.40$, $x_{right} = 0.938$.

## 4. Flow around the ellipse under angle of attack

The parameter $\Delta t$ for a separated flow around the ellipse under angle of attack should be determined from the condition of the separation points velocities equality, which are found from equations (3.3) and (3.4).

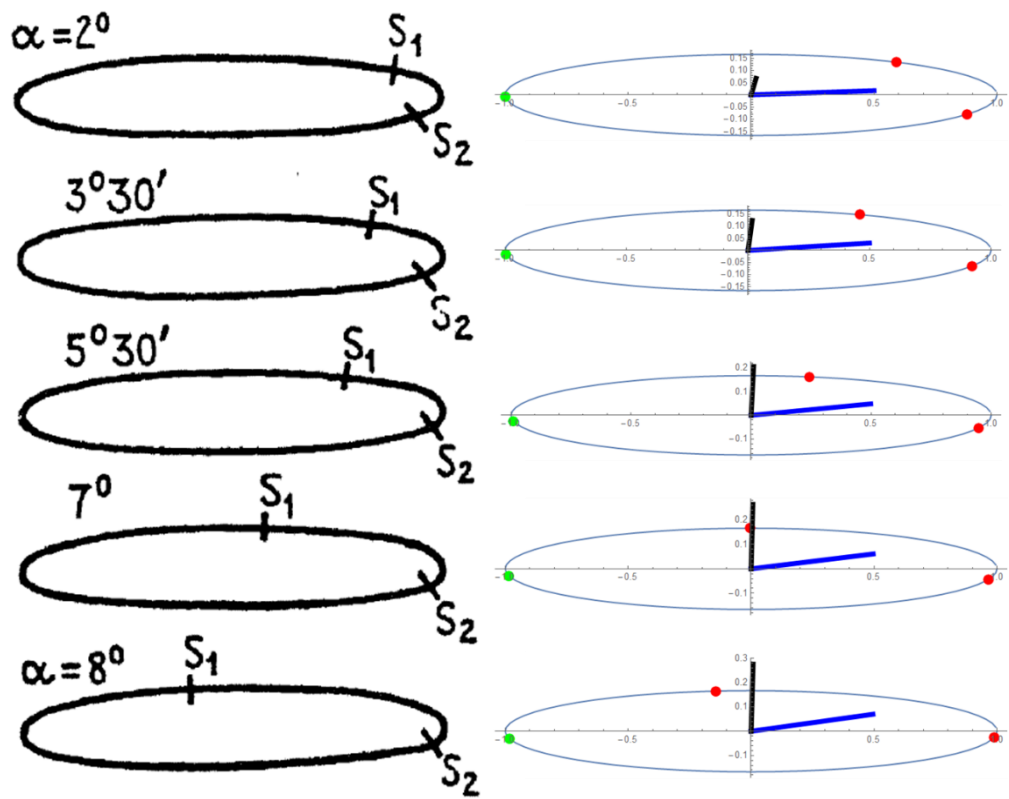

Fig 4. Separation points in case of the ellipse axis ratio 6:1 under different angles of attack.

Fig. 4 shows the breakaway points for the flow around an ellipse with an axis ratio of 6: 1 at angles of attack: 2, 3.5, 5.5, 7 and 8 degrees. On the left, the calculation was performed by the Karman-Pohlhausen method [6] and on the right by the described earlier method. Blue line is velocity direction and black line is force direction.

## 5. Circulation and lift force

Lift force $Y$ is expressed in terms of circulation as follows $Y = \rho \Gamma v_\infty$, where circulation is expressed in terms of a parameter $\Delta t$ by formula (3.2). Fig. 5 shows a comparison of calculations of the dependence of the lift coefficient on the angle of attack.

$$C_y = \frac{Y}{\rho v_\infty^2 / 2} = 4\pi(a_{el} + b_{el})\sin(\theta + \Delta t)$$

For the ellipse $a_{el} = 0.54$, $b_{el} = 0.1$, according to the above method, the calculations are given for the flow under angles of attack: 2, 3.5, 5.5, 7 and 8 degrees (red dots). The solid line is given in the monograph [6].

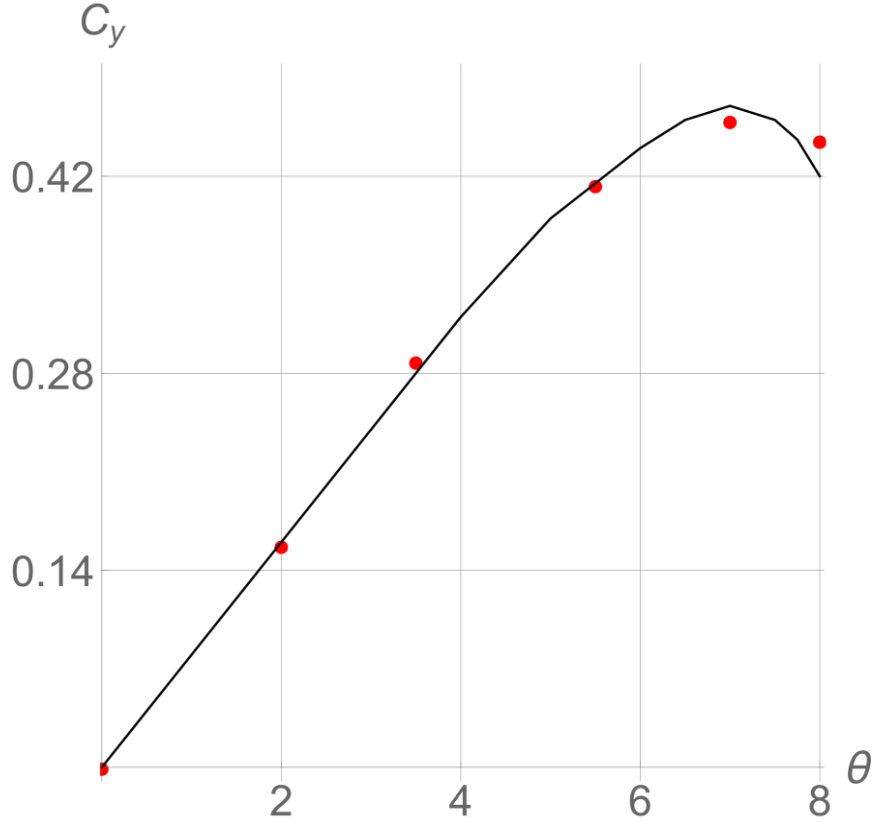

Fig. 5. Comparison of calculations of the dependence of lift force on the angle of attack.

## 5. Interpolation formula for lift force coefficient

Calculation of lift force coefficient $C_y = \dfrac{Y}{\rho v_\infty^2 a_{el}/2} = 4\pi(1 + \dfrac{b_{el}}{a_{el}})\sin(\theta + \Delta t)$ for different attack angles $\theta$: 0, 1, ... 9 degrees and for different axis ratio $\dfrac{b_{el}}{a_{el}}$: 0.1, 0.12, … 0.26 is produced. Approximation is found from the results:

$$C_y \approx \left(-0.0056\theta_{deg}^3 + 0.0178\theta_{deg}^2 + 0.4237\theta_{deg}\right)\frac{b_{el}}{a_{el}} + \left(-0.002\theta_{deg}^3 + 0.0068\theta_{deg}^2 + 0.2147\theta_{deg}\right) \quad (5.1)$$

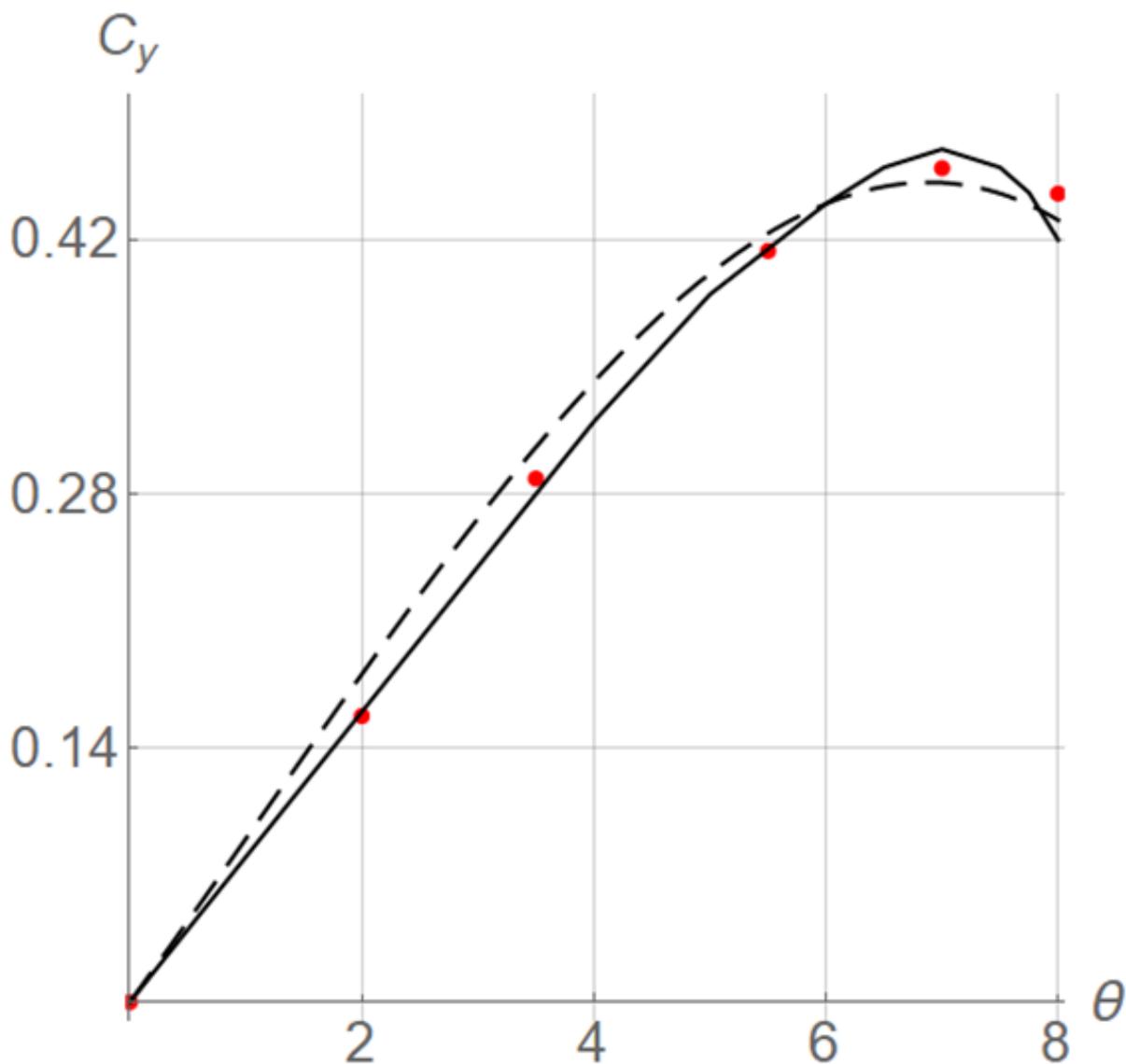

Fig. 6. Comparison of calculations of the dependence of lift force on the angle of attack. Approximation - dash line.

In the Fig. 6 showed approximation dash line by the formula (5.1). The discrepancy in the surrounding area of maximum is explained by the insufficient number of data for approximation.

## 6. Acknowledgements

The reported investigation was funded by RFBR and NSFC, project number 21-57-53019 and was carried out within the state assignment of FASO of Russia (state registration № AAAA-A20-120011690138-6).